\documentclass[letter]{aa} 

\usepackage{graphicx}
\usepackage{txfonts}
\usepackage{multirow} 
\newcommand{\Msun}{M$_{\odot}$}

\newcommand{\gppr}{\stackrel{>}{\scriptstyle \sim}}
\newcommand{\gappr}{\raisebox{-0.4ex}{$\gppr$}}
\newcommand{\lppr}{\stackrel{<}{\scriptstyle \sim}}
\newcommand{\lappr}{\raisebox{-0.4ex}{$\lppr$}}
\definecolor{smalt(darkpowderblue)}{rgb}{0.0, 0.2, 0.6}
\definecolor{forestgreen(traditional)}{rgb}{0.0, 0.5, 0.0}

\defcitealias{RVJ}{RVJ}

\defcitealias{MD17}{MD17}

\newcommand{\cv}{CV}
\newcommand{\cvs}{CVs}

\newcommand{\pcebs}{post-CE binaries}

\newcommand{\mesa}{MESA}

\usepackage{color}

\begin{document}

   \title{Period bouncers as detached magnetic cataclysmic variables} 


   \author{
          Matthias R. Schreiber\inst{1.2}
          \and
          Diogo Belloni
          \inst{1}
          \and
          Jan van Roestel\inst{3}
          }

    \authorrunning{Schreiber \& Belloni}

  \institute{Departamento de F\'isica, Universidad T\'ecnica Federico Santa Mar\'ia, Av. España 1680, Valpara\'iso, Chile\\
              \email{matthias.schreiber@usm.cl, diogobellonizorzi@gmail.com}
         \and
             Millennium Nucleus for Planet Formation, Valpara{\'i}so, Chile
        \and
            Anton Pannekoek Institute for Astronomy, University of Amsterdam, NL-1090 GE Amsterdam, the Netherlands
             }

   \date{}

 
  \abstract
   {The general prediction that more than half of all cataclysmic variables (\cvs)~have evolved past the period minimum is in strong disagreement with observational surveys, which show that the relative number of these objects is just a few per cent.} 
   {Here, we investigate whether a large number of post-period minimum \cvs~could detach because of the appearance of a strong white dwarf magnetic field potentially generated by a rotation- and crystallization-driven dynamo.}
   {We used the \mesa~code to calculate evolutionary tracks of \cvs~incorporating the spin evolution and cooling as well as compressional heating of the white dwarf. If the conditions for the dynamo were met, we assumed that the emerging magnetic field of the white dwarf connects to that of the companion star and incorporated the corresponding synchronization torque, which transfers spin angular momentum to the orbit.}
   {We find that for \cvs~with donor masses exceeding $\sim0.04$\Msun,~magnetic fields are generated mostly if the white dwarfs start to crystallize before the onset of mass transfer. 
   It is possible that a few white dwarf magnetic fields are generated in the period gap. 
   For the remaining \cvs,~the conditions for the dynamo to work are met beyond the period minimum, when the accretion rate decreased significantly. Synchronization torques cause these systems to detach for several gigayears even if the magnetic field strength of the white dwarf is just one MG. }
   {If the rotation- and crystallization-driven
dynamo ---which is currently the only mechanism that can explain several observational facts related to magnetism in \cvs~and their progenitors--- or a similar temperature-dependent mechanism is responsible for the generation of magnetic field in white dwarfs, most \cvs~that have evolved beyond the period minimum must detach for several gigayears at some point. This reduces the predicted number of semi-detached period bouncers by up to $\sim60-80$ per cent. }

   \maketitle
%

\section{Introduction}

According to the standard theory for the formation and evolution of cataclysmic variables (CVs, \citealt{belloni+schreiber23-1,kniggeetal11-1}), and independent of the details of simulating CV evolution, a large number of all \cvs~should already have passed the period minimum, that is, the population of \cvs~should be dominated by systems evolving toward longer orbital periods consisting of a cool white dwarf with a brown dwarf companion.  
The predicted fraction of these so-called period bouncers among 
\cvs~was estimated decades ago  to be 70 per cent \citep{kolb93-1}, a value roughly in agreement with more recent calculations, which predict $38-60$ per cent \citep{goliasch+nelson15-1} and $\sim75$ per cent \citep{bellonietal18-1}.

However, despite significant observational efforts, very few period bouncers have so far been identified. 
\citet{patterson11-1} 
heard some murmurs from the period bounce
by identifying 22 candidates
among the \cvs~known at that time. 
More recently, \citet{palaetal20-1} found the fraction of period bouncers to be 
$7-14$ per cent in a volume-limited sample of \cvs. 
Dedicated surveys for \cvs~in the data releases of the Sloan Digital Sky Survey (SDSS) confirm that 
the fraction of period bouncers does not exceed a few per cent \citep{inightetal23-1,inightetal23-2}. 
This large discrepancy between theoretical prediction and observations 
might be the most important problem in our understanding of CV evolution. 

A potentially important ingredient in CV evolution ---that was ignored in binary population models of \cvs~for decades--- is the possible impact of white dwarf magnetic fields, despite the fact that more than one-third of all \cvs~in volume-limited samples host a magnetic white dwarf \citep{palaetal20-1}. 
Observations of both single white dwarfs \citep{bagnulo+landstreet21-1} and detached white dwarf binaries with M-dwarf companions
\citep{parsonsetal21-1}
clearly show that the vast majority of white dwarf magnetic fields appear when the white dwarf has significantly cooled, that is, either the magnetic field is buried for several gigayears (Gyr) after the white dwarf formation, or the conditions for generating a magnetic field are met only in cool white dwarfs. 

Based on these observational facts and the early work by \citet{isernetal17-1}, 
we developed the idea of a rotation- and crystallization-driven 
magnetic dynamo that generates strong white dwarf magnetic fields during CV evolution \citep{schreiberetal21-1}. 
According to this model, white dwarfs that crystallize before mass-transfer starts generate a strong magnetic field as soon as accretion has spun up the white dwarf. If the emerging magnetic field connects with that of the donor star, spin angular momentum is transferred to the orbit, which causes the separation of both stars to increase and the CV to convert into a detached binary. Loss of angular momentum  through gravitational radiation and/or reduced magnetic braking \citep{bellonietal20-1} then slowly shrinks the orbit until the donor fills its Roche lobe and the system becomes a magnetic CV.

While the assumption of magnetic field generation through the dynamo is rather speculative (as other ideas for the origin of magnetic fields in white dwarfs are), the described evolutionary scenario can reproduce a number of observations that otherwise remain inexplicable: (i) the fact that so far all observed 
detached close binaries containing a strongly magnetic white dwarf are close to Roche-lobe filling and typically have periods of between $3$ and $5$\,hr; (ii) the existence of the radio pulsing white dwarf binaries AR\,Sco \citep{marshetal16-1} and eRASSU\,J191213.9--441044 \citep{pelisolietal23-1}; (iii) the small number of high-accretion-rate magnetic CVs in globular clusters \citep{bellonietal21-1}; and (iv) the observed magnetic fields in detached double white dwarfs \citep{schreiberetal22-1}. 

Here we focus on magnetic field generation and angular momentum transfer for systems that evolve toward and beyond the period minimum as nonmagnetic \cvs. In these systems, the mass-transfer rate becomes small enough to allow the core of the white dwarf to crystallize, and according to the dynamo idea, the white dwarf should develop a magnetic field. The corresponding synchronization torques should then cause the system to detach for several Gyr, thereby reducing the number of predicted period bouncers significantly. 

\section{The model}

We used the \mesa~code \citep[][r15140]{Paxton2011, Paxton2013, Paxton2015, Paxton2018, Paxton2019,jermynetal23-1} to compute the evolution of \cvs~as in \citet{schreiberetal21-1}. The initial conditions for all our simulations are detached \pcebs~with different initial donor star and white dwarf masses and orbital periods. 

We assumed angular momentum loss through magnetic braking according to \citet{rappaportetal83-1} and through gravitational radiation, and that magnetic braking gets disrupted when the donor star becomes fully convective. 
We used normalization factors to calibrate the strength of angular momentum loss. 
We also considered the empirical prescription for consequential angular momentum loss found by \citet{schreiberetal16-1}, which is required to bring the observed and predicted white dwarf mass distributions  into
agreement. 

Angular momentum accretion and spin-up of the white dwarf during mass transfer was calculated as in \citet{schreiberetal21-1}, that is, largely following \citet{kingetal91-1}. The potential connection between the emerging white dwarf magnetic field and that of the donor star, as well as the resulting angular momentum transfer of spin angular momentum to orbital angular momentum, were 
also treated as in \citet{schreiberetal21-1}. 

In line with \citet{ginzburgetal22-1}, we assumed the critical spin period to be one hour, which is significantly longer than the values assumed by \citet{schreiberetal21-1}. We also updated our model by incorporating a diffusion timescale that reflects the time the field requires to reach the surface of the white dwarf after the conditions for magnetic field generation are met in the core \citep{ginzburgetal22-1}.

Crystallization depends on the core temperature of the white dwarf, which is therefore of fundamental importance for magnetic field generation in our model. 
We assumed the critical temperature for the dynamo to work to be the temperature corresponding to the onset of crystallization, as provided by the thick hydrogen atmosphere models of \citet{bedardetal20-1}. 
The core temperature of cooling single white dwarfs can be reliably determined from these evolutionary sequences. 

The most important update compared to our previous modeling of \cv~evolution including magnetic field generation through the rotation-  and crystallization-driven dynamo concerns the heating of the white dwarf core through accretion. Accretion can slowly (on a timescale of a few 100 million years) heat up the core until an equilibrium at the base of the radiative layer is reached. 

Incorporating the full details of cooling and heating of accreting white dwarfs in our simulations is beyond the scope of this paper. However, based on published results \citep[][]{townsley+bildsten04-1, epelstainetal07-1, townsley+gaensicke09-1}, we developed a simple prescription that should cover the main physics.  
Briefly, we used the cooling temperature for nonaccreting white dwarfs while for accreting white dwarfs we slowly increased the core temperature until the equilibrium core temperature (governed by the mass accretion rate) was reached. 

We are aware that our modeling approach is based on several assumptions, such as the stellar field strengths, the accretion efficiency of angular momentum, and the synchronization or diffusion timescales. In addition, our prescription for determining the white dwarf core temperature during accretion is clearly a simplified representation of reality. 
However, the general idea of the scenario is independent of the exact values and procedures that we assumed in our modeling.  
More details can be found in \citet{schreiberetal21-1} and Appendix\,\ref{app_mod}. 
The code we used to calculate the evolutionary tracks can be found here: https://zenodo.org/records/10008722.

\section{Evolutionary pathways}

\begin{figure}
\begin{center}
\includegraphics[width=0.99\linewidth]{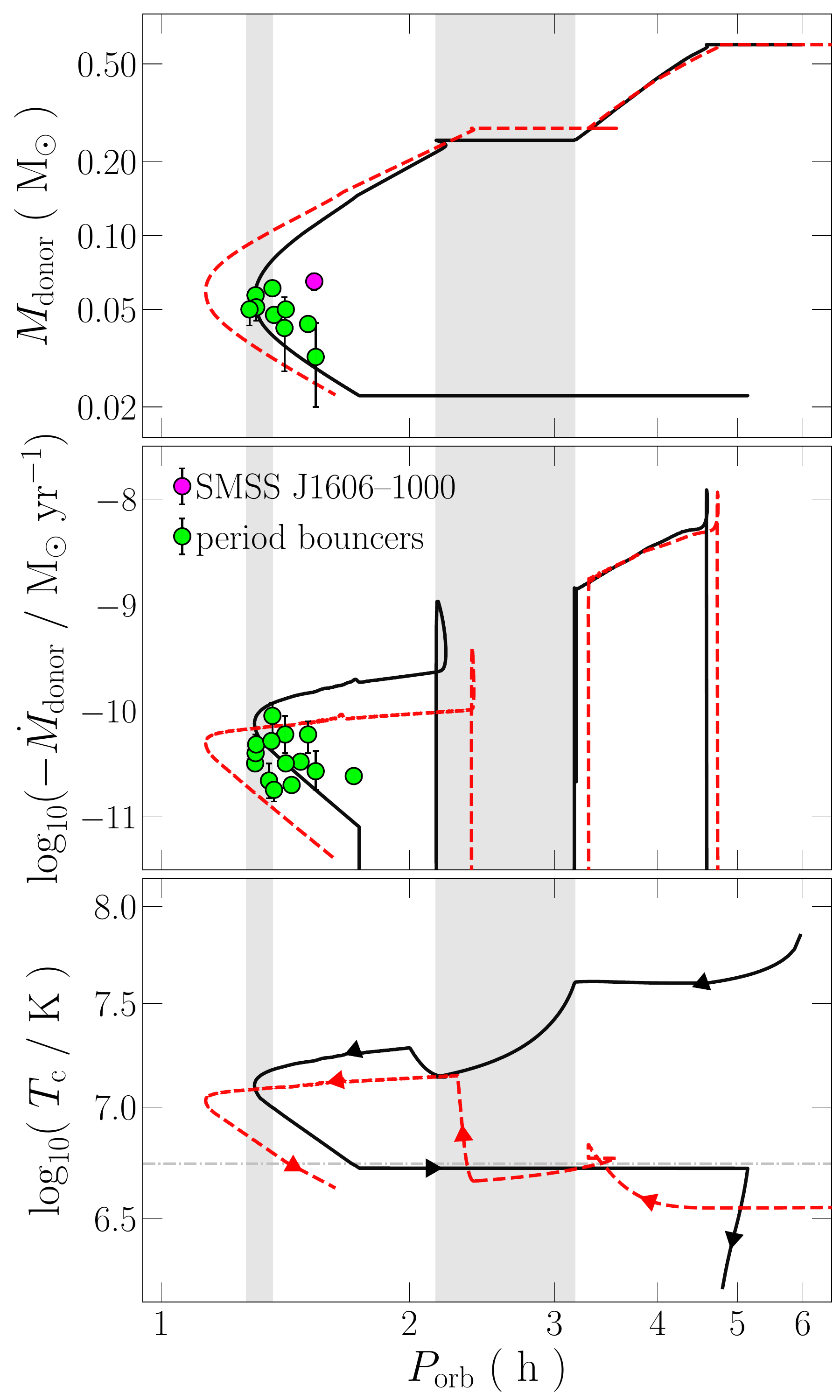}
\end{center}
\caption{Evolutionary tracks for two systems starting mass transfer above the gap. The initial parameters are $M_{\rm WD} = 0.8$\,\Msun, $M_2=0.6$\,\Msun, and $P_{\mathrm{orb}}=0.25$ (black) and $1.0$ days (red). The horizontal gray line in the bottom panel corresponds to $\log(Tc)=6.742366$, which is the core temperature at the onset of crystallization. The vertical lines refer to the period minimum of $P_{\mathrm{min}} = 76-82$ minutes, and the period gap ($2.15-3.18$\,hours). While the system represented by the red track detaches above the gap because the white dwarf crystallizes before mass transfer starts, \cvs~that start mass transfer with a noncrystallizing white dwarf detach after they have passed the period minimum (black track). We note that the period minimum is at shorter periods for systems that previously generated a white dwarf magnetic field. This is because magnetic braking is reduced as some of the open field lines of the donor star are affected by the white dwarf magnetic field.}
\label{Fig_agap}
\end{figure}

Figure\,\ref{Fig_agap} shows two evolutionary tracks for \cvs~that start mass transfer above the orbital period gap, together with observed period bouncers and the detached magnetic white dwarf with a brown dwarf companion SMSSJ~160639.78$-$100010.7 (see Appendix\,\ref{app_obs} for more details on the observed systems). The red dashed lines represent a case similar to those described in \citet{schreiberetal21-1}, that is, mass transfer starts when the white dwarf core has already started to crystallize. The magnetic field is generated in the core of the white dwarf when the critical rotation rate is reached. 
The field then diffuses outward and 
when it emerges at the surface, it can connect with the field of the secondary star and synchronization torques lead to the transfer of spin angular momentum to the orbit, and the binary becomes a detached system. The model predicts the detached binary to first resemble AR\,Sco \citep{marshetal16-1} and later, when spin and orbit are synchronized, to become a pre-polar \citep{parsonsetal21-1}. 

The black solid line corresponds to the case of a white dwarf that is still far from crystallizing when mass transfer starts. This prevents the system from generating a magnetic field above the gap because accretion keeps the core temperature above the crystallization limit. The system detaches at the upper 
edge of the orbital period gap as predicted by standard \cv~evolution theory. While crossing the gap, the white dwarf cools but not enough to start crystallizing because with residual magnetic braking, the evolution through the gap takes only $\sim300$\,Myr. 
When mass transfer resumes below the gap, accretion determines the core temperature. 
As the mass-accretion rate decreases significantly when the donor converts into a brown dwarf and the system passes the orbital period minimum, the core temperature of the white dwarf inevitably reaches the crystallization limit at some point. This causes the dynamo to generate a magnetic field which, after it has diffused to the white dwarf surface, connects with that of the donor and causes the secondary to detach from its Roche lobe. 

The generated detached phase takes several Hubble times as the spin period has reached very small values because of the large amount of accreted angular momentum and because the remaining angular-momentum-loss mechanism is gravitational radiation (residual magnetic braking is suppressed through the connection of both magnetic fields; e.g. \citealt[][]{bellonietal20-1}). 
The field strength needed to connect with the secondary star is much smaller than above the gap because the binary separation and mass-transfer rate are much smaller \citep[see][their eq. 9]{schreiberetal21-1}. As little as $\sim1$\,MG is sufficient to generate a long detached phase. 

In full analogy to what happens in the case where strong fields are generated in the 3-5 hour orbital period range, 
the system appears first as a radio-pulsating fast-spinning white dwarf with a brown dwarf companion, before it synchronizes and becomes a pre-polar. 
Given that the spin angular momentum is typically high enough for the system to quickly evolve to periods of up to five hours, the age of the Galaxy is not sufficient for these detached systems to evolve back into a semi-detached configuration. 

The tracks discussed above are representative of the majority of \cvs,~as most are born above the gap. If the core crystallizes before the onset of mass transfer, the magnetic field appears above or in the period gap. If, on the other hand, the white dwarf does not crystallize before the onset of mass transfer, it will only do so when the mass-transfer rate significantly drops after the system has passed the period minimum.  

\begin{figure}
\begin{center}
\includegraphics[width=0.99\linewidth]{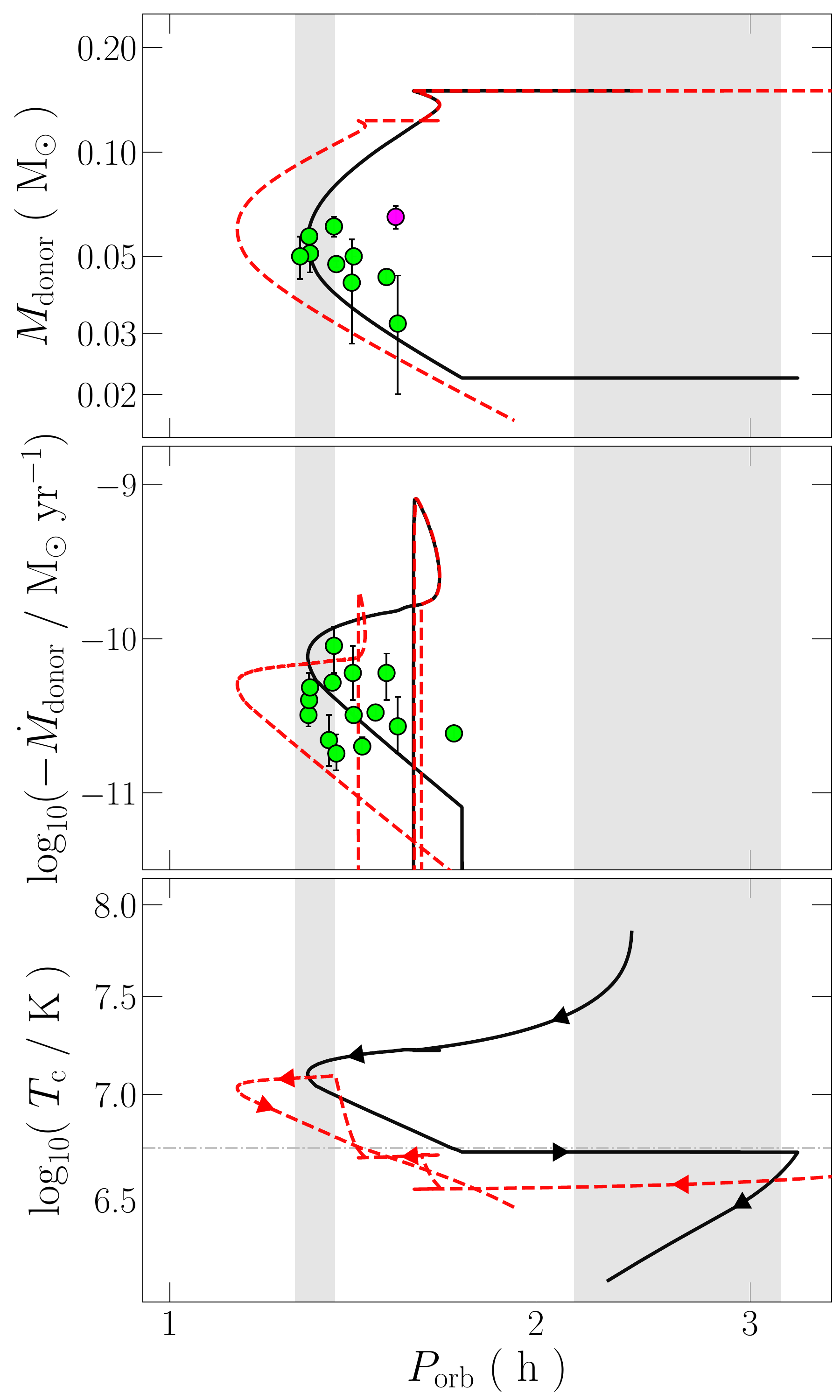}
\end{center}
\caption{Examples of evolutionary tracks for \cvs~starting mass transfer below the gap. The initial donor mass is assumed to be ${M_2=0.15}$\,\Msun~and with initial post-common-envelope periods of $0.10$ (black) and $0.25$ days (red). If a \cv~is born with a crystallizing white dwarf, the magnetic field is generated before the period minimum is reached and the system detaches for a relatively short period of time ($\sim130$\, Myr) and then evolves as a magnetic CV through the period minimum (red dashed track). If the white dwarf does not start to crystallize before the onset of mass transfer, the binary detaches after it has passed the period minimum when the accretion rate became low enough to allow the white dwarf to crystallize (black solid track). }
\label{Fig_bgap}
\end{figure}

The situation is similar for \cvs~that start mass transfer in or below the gap. Figure\,\ref{Fig_bgap} shows two tracks that describe the possible evolutionary pathways for such objects. 
The red dashed line corresponds to an evolution where the core of the white dwarf crystallizes before the onset of mass transfer, which causes the field to appear when the rotation criterion is met and the field has had time to diffuse to the white dwarf surface. This causes a relatively short ($\sim300$\,Myr) detached phase that occurs at a period similar to that of the onset of mass transfer (and before the period minimum is reached). 

If the core does not crystallize before the onset of mass transfer (black line in Fig.\,\ref{Fig_bgap}), 
compressional heating of the white dwarf core during the mass transfer phase prevent 
crystallization until the mass accretion rate drops, which occurs after the binary has passed the period minimum. Compared to the case of \cvs~that started mass transfer above the gap with a noncrystallizing white dwarf, 
the maximum period reached during the detached phase is shorter because less angular momentum has been accreted, but it still takes longer than a Hubble time to evolve back to mass transfer. 

While the exact evolution of a given system depends on the masses and orbital period of the detached post-common-envelope binary, virtually all \cvs~are predicted to evolve according to one of the four scenarios described above. The exceptions are \cvs~that become magnetic in the period gap. This might occur when the core of the white dwarf cools substantially before mass transfer but does not crystallize and accretion 
does not fully heat up the white dwarf core. To match all these conditions, some fine-tuning is required and therefore we expect relatively few systems to evolve in this way.

\section{Predictions}

\citet{schreiberetal21-1} invented the rotation- and crystallization-driven dynamo to explain the existence of detached post-common-envelope binaries with cool magnetic white dwarfs close to Roche-lobe filling in the period range of 3 to 5\,hr, the radio pulsing white dwarf binary AR\,Sco, and the fact that a large number of \cvs~contain a magnetic white dwarf. We here focus on the predictions of the model with respect to the evolution and magnetic field generation for \cvs~below the orbital period gap.

The relative number of \cvs~that are predicted to generate a magnetic field before the period minimum is reached, that is, systems that either crystallize before mass transfer starts or in the period gap, depends on the initial mass-ratio distribution, common envelope efficiency, and the strengths of magnetic braking. 
Performing detailed binary population synthesis is beyond the scope of this paper. 
However, assuming that a set of parameters for common envelope evolution, magnetic braking, and the initial mass ratio distribution exists that can explain the measured 
fraction of 
magnetic \cvs, which is in the range of $20-40$ per cent \citep{palaetal20-1,inightetal23-1,inightetal23-2}, it is straightforward to make testable predictions.

If the dynamo is fully responsible for magnetic field generation in \cvs, $\sim20-40$ per cent of all 
\cvs~either hosted a crystallizing white dwarf at the onset of mass transfer or the white dwarf crystallized in the period gap. These systems will pass the period minimum as magnetic \cvs. 
The remaining $\sim60-80$ per cent of all \cvs~evolve toward the period minimum as nonmagnetic \cvs. 
Only after evolving through the period minimum can the core crystallize which causes the magnetic field to emerge and the binary to become a detached binary consisting of a magnetic white dwarf ($B\geq1$\,MG) with a brown dwarf companion. 
This reduces the fraction of expected period bouncers from $38-75$ per cent \citep{kolb93-1,goliasch+nelson15-1,bellonietal18-1} to $\sim8-30$ per cent if we ignore the post-period-minimum evolution before the detached phase. 
This rough estimate represents an upper limit. If nonmagnetic \cvs~spend a significant time span as accreting \cvs~after having passed the period minimum and before the accretion rate sufficiently drops for the core to crystallize, the reduction of period bouncers might be 
significantly smaller than $60-80$ per cent. 
How long a given system spends as an accreting \cv~after having passed the period minimum depends 
on the crystallization temperature (which depends on the white dwarf mass) and the strength of angular momentum loss at this stage, which is not well known.

While detailed binary population synthesis is required to 
provide the exact numbers, we conclude that the dynamo, which was invented to explain the magnetic nature of \cvs~and detached systems with periods longer than $3$\,hr, may bring the predicted relative numbers of period bouncers into agreement with the currently available observational constraints. 
The trade-off of this result is that the model predicts a large number of cool magnetic white dwarfs with brown dwarf companions, with a predicted space density similar to that of \cvs~with stellar companions.  
These systems might be difficult to detect because of the faintness of both stellar components and the absence of accretion.  
So far, there is only one known magnetic white dwarf that has a brown dwarf companion in a clearly detached binary that could be a detached CV \citep{kawkaetal21-1}. 

\section{Conclusion}

If a rotation- and crystallization-driven dynamo generates the magnetic field and the explanation for the existence of AR\,Sco and the detached magnetic white dwarf binaries in the orbital period range of $3-5$\,hr presented by \citet{schreiberetal21-1} is correct, as a natural consequence, a large fraction of period bouncers must detach as soon as the accretion rate drops sufficiently and the white dwarf starts to crystallize. This may in principle explain the low number of period bouncers identified so far. 

Even if the dynamo idea is found to be unfeasible, for example because
the energy carried as kinetic energy in the convection zone is found to be insufficient 
to drive an efficient dynamo
\citep{fuentesetal23-1}, the evolutionary scenario outlined in \citet{schreiberetal21-1} and in this letter remains plausible as long as the magnetic field appearance in white dwarfs depends on their core temperature.  
There is indeed overwhelming evidence for the increased occurrence of magnetic fields in cool white dwarfs 
\citep{parsonsetal21-1,bagnulo+landstreet21-1}. 

Unequivocal proof for a temperature-dependent magnetic field generation in white dwarfs would be provided by the detection of a large number of detached binaries consisting of a cool ($\lappr10\,000$\,K) magnetic ($\gappr1$\,MG) white dwarf with a brown dwarf companion and orbital periods of between 80 minutes and a few hours. At present, just one such system is known \citep{kawkaetal21-1}.

\begin{acknowledgements}

We are very grateful to the Kavli Institute for Theoretical Physics (KITP) for hosting the program ``White Dwarfs as Probes of the Evolution of Planets, Stars, the Milky Way and the Expanding Universe''.
This research was supported in part by the National Science Foundation under Grant No. NSF PHY-1748958
and by the Munich Institute for Astro-, Particle and BioPhysics (MIAPbP) which is funded by the Deutsche Forschungsgemeinschaft (DFG, German Research Foundation) under Germany's Excellence Strategy – EXC-2094 – 390783311.
MRS was supported from FONDECYT (grant number 1221059) and ANID, – Millennium Science Initiative Program – NCN19\_171.
DB acknowledges financial support from FONDECYT (grant number 3220167).

\end{acknowledgements}

%
%

%

\begin{appendix}

\section{Modeling details}
\label{app_mod}
The equations describing most of the ingredients of the model, such as the white dwarf spin-up and orbital angular momentum loss, are given in \citet{schreiberetal21-1}. We report all parameters and revised assumptions below. 

\subsection{MESA}

We used the \mesa~code \citep[][r15140]{Paxton2011, Paxton2013, Paxton2015, Paxton2018, Paxton2019,jermynetal23-1} to compute the evolution of \cvs~and their progenitors.
The \mesa~equation of state is a blend of the OPAL \citep{Rogers2002}, SCVH \citep{Saumon1995}, FreeEOS \citep{Irwin2004}, HELM \citep{Timmes2000}, PC \citep{Potekhin2010} and Skye \citep{Jermyn_2021} equations of state.
Nuclear reaction rates are a combination of rates from NACRE \citep{Angulo1999}, JINA REACLIB \citep{Cyburt2010}, plus additional tabulated weak reaction rates \citep{Fuller1985, Oda1994,Langanke2000}.
Screening is included via the prescription of \citet{Chugunov2007} and thermal neutrino loss rates are from \citet{Itoh1996}.
Electron conduction opacities are from \citet{Cassisi2007} and radiative opacities are primarily from OPAL \citep{Iglesias1993,Iglesias1996}, with the high-temperature Compton-scattering-dominated regime calculated using the equations of \citet{Buchler1976}.

\subsection{Spin-up of the white dwarf}

The spin up of accreting white dwarfs was calculated 
as in \citet[][their eq.\,1]{schreiberetal21-1} assuming an initial spin period of one year 
and the maximum spin-up efficiency of $\alpha=1$. 
The angular momentum loss parameter of the white dwarf due to nova explosions was set to $\eta=0.833$ which corresponds to a slightly nonspherical ejection. 
The critical spin period for the generation of a magnetic field was set to one hour. We assumed the white dwarf mass to be constant during \cv-evolution, that is, we assume the accreted mass 
to be ejected during nova eruptions. 

\subsection{Orbital angular momentum loss}

For secondary stars with a radiative core and convective envelope, we assumed angular momentum loss through magnetic braking according to \citet[][with $\gamma=3$]{rappaportetal83-1} with a normalization factor of $0.6$ \citep{kniggeetal11-1}. 
In case the magnetic field of the white dwarf connects with that of the secondary star, we assumed that magnetic braking is reduced as described in \citet{webbink+wickramasinghe02-1} and \citet{bellonietal20-1}. 

Angular momentum loss through gravitational radiation was calculated 
using the weak-field approximation of general relativity as in \citet{hurleyetal02-1}. 
We assumed residual magnetic braking for fully convective donor stars and approximated it by multiplying 
angular momentum loss through gravitational radiation by $1.5$ which provides accretion rates below the gap in agreement with the observations \citep{kniggeetal11-1}. We assumed the empirical consequential angular momentum loss prescription of \citet[][their eq.\,5]{schreiberetal16-1} with a normalization factor of $C=0.35$.

\subsection{Cooling and heating of the white dwarf} 

We follow the cooling of nonaccreting white dwarfs using the models by \citet{bedardetal20-1}. 
%
Incorporating the full details of cooling and heating of accreting white dwarfs in our simulations is beyond the scope of this paper. 
We instead incorporated a simple prescription based on published results  \citep{townsley+bildsten04-1,epelstainetal07-1,townsley+gaensicke09-1}. 

In cases where the cooling temperature of the white dwarf at the onset of mass transfer is below the temperature generated by compressional heating  \citep[][their equation 1]{townsley+gaensicke09-1}, 
the white dwarf core temperature is heated on timescales of a few  hundred million years until the equilibrium luminosity ${L_{\mathrm{eq}}\simeq\langle\dot{M}\rangle T_{\mathrm{c}}/\mu m_{\mathrm{p}} = 4\pi R^2\sigma T_{\mathrm{eff}}^4}$ is reached. Here, $T_{\mathrm{c}}$ is the core temperature assumed to be equal to that at the base of the radiative layers,
$R$ is the white dwarf radius, $\mu$ is the mean molecular
weight of the accreted material, $m_{\mathrm{p}}$ is the mass of the proton,
and $\sigma$ is the Stefan–Boltzmann constant \citep[see][for more details]{townsley+gaensicke09-1}. 

Throughout this work we assume that the timescale required to reach the equilibrium temperature is $200$\,Myr. 
When mass transfer starts and the compressional heating temperature exceeds the cooling temperature of the white dwarf, we slowly increase the core temperature interpolating between the cooling temperature and the equilibrium temperature using a cubic equation. The latter makes sure that the increase in temperature is slow at the beginning \citep{epelstainetal07-1}. If mass transfer stops, the white dwarf temperature is assumed to follow again the cooling models from \citet{bedardetal20-1} but starting from the core temperature reached during accretion.  

In the case where the cooling temperature of the white dwarf is hotter than the temperature derived from compressional heating, the cooling temperature of course should represent the true white dwarf core temperature. 
In this case, we therefore assume the usual cooling of a nonaccreting white dwarf to determine the timescale on which the equilibrium between compressional heating and core temperature is reached. 

\subsection{Dynamo conditions and synchronization}

We assumed the generation of a magnetic field to require the core temperature of the white dwarf to have reached the onset of crystallization according to the cooling models by \citet{bedardetal20-1} and a spin period shorter than one hour. 
Once the conditions for the dynamo to work are met, we assume a delay for the appearance of the magnetic field on the surface of the white dwarf. 
This timescale depends on the mass and the temperature of the white dwarf \citep[][their fig. 3]{ginzburgetal22-1}. Here, we only show tracks for white dwarfs of the canonical mass for white dwarfs in \cvs~($0.8$\,\Msun, \citealt[][]{zorotovicetal11-1}) and assume a diffusion timescale of $10^8$\,yr.

We assumed a magnetic field of $1$\,kG for the donor star independent of its mass and the generated white dwarf magnetic field to be of $60$\,MG. 
The synchronization timescale as defined in \citet{schreiberetal21-1} was set to $1$\,Myr. 

\subsection{Other parameters}

We assumed a maximum evolutionary time of $10$\,Gyr. We start all our simulations with a detached post-common-envelope binary assuming 
that the formation of the white dwarf takes $300$\,Myr. 
We calculated the evolution for a white dwarf mass of $0.8$\,\Msun~and donor masses of $0.15$ and $0.6$\,\Msun. The initial orbital periods of the post common envelope binary were assumed to be between $0.1-1$\,days.

\section{Observed period bouncers}
\label{app_obs}

In order to be able to compare our evolutionary tracks with observed systems, we compiled a list of observed period bouncers (Table\,\ref{tab:bouncers}). We only included systems with measured orbital period as well as donor mass and/or accretion rate.  

Three out of $20$ systems contain a magnetic white dwarf which corresponds to a fraction of $15\pm7$ per cent in rough agreement with the number of magnetic \cvs~in the SDSS sample \citep{inightetal23-1,inightetal23-2} but smaller than in the volume limited  150\,pc sample \citep{palaetal20-1}. 

We found one system that could perhaps be a detached magnetic \cv: SMSSJ~160639.78$-$100010.7 consists of a magnetic white dwarf with a brown dwarf companion orbiting their common center of mass in $92$ minutes \citep{kawkaetal21-1}. While for the measured white dwarf mass, the temperature seems to be slightly too hot for the white dwarf to be crystallizing, the system could be explained by a \cv~with a white dwarf that crystallized before the onset of mass transfer. In this scenario, the mass transfer must have started below the gap with a low-mass M-dwarf companion ($\lappr0.1$\,\Msun). 
Two promising candidate detached magnetic period bouncers are ZTF\,J014635.73+491443.1 and SDSS\,J121209.31+013627.7 \citep{hakalaetal21-1, guidryetal21-1}. We do not list these candidates in Table\,\ref{tab:bouncers} because of uncertainties related to the measured period or the nature of the donor star. Interestingly, ZTF\,J014635.73+491443.1 is nearby (56\,pc), which might indicate that such systems are not rare.

\begin{table*}
\centering
\caption{Properties of confirmed and candidate period bouncers.} 
\label{tab:bouncers}
\setlength\tabcolsep{10pt} 
\renewcommand{\arraystretch}{1.5} 
\begin{tabular}{lccccccr}
\hline
\hline
\vspace{-0.1cm}
\multirow{2}{*}{System} & $P_{\rm orb}$ & $\langle\dot{M}_{\rm d}\rangle$ & $M_{\rm d}$ & ref. \\
                        & (min)         & ($10^{-10}$~\Msun~yr$^{-1}$)    & (\Msun)     & \\
\hline
SDSS~J143544.02$+$233638.7             &  $78.00$ & $0.32^{+0.08}_{-0.05}$    &                   & 1, 2 \\
SDSS~J143317.78$+$101123.3             &  $78.10$ & $0.40^{+0.20}_{-0.10}$    & $0.0571\pm0.001$  & 1, 2 \\
V455~And                               &  $81.08$ & $0.22^{+0.10}_{-0.07}$    &                   & 1, 2\\
SDSS~J150137.22$+$550123.3         &  $81.85$ & $0.90\pm0.30$             & $0.061\pm0.004$   & 1, 2\\
SDSS~J103533.02$+$055158.4         &  $82.22$ & $0.18^{+0.06}_{-0.04}$    & $0.0475\pm0.001$  & 3, 2\\
SDSS~J150240.98$+$333423.9         &  $84.83$ & $0.6^{+0.3}_{-0.2}$       &                   & 4 \\
EG~Cnc                             &  $86.36$ & $0.2^{+0.03}_{-0.02}$     &                   & 4 \\
1RXS~J105010.8$-$140431            &  $88.56$ & $0.332^{+0.028}_{-0.018}$ &                   & 4 \\
QZ~Lib                             &  $92.36$ & $0.27^{+0.15}_{-0.09}$    & $0.032\pm0.012$   & 4 \\     
GD~552                             & $102.73$ & $0.243^{+0.023}_{-0.018}$ &                   & 5, 6 \\
BW~Scl                             &  $78.23$ & $0.483^{+0.014}_{-0.012}$ & $0.051\pm0.006$   & 7, 6 \\ 
WZ~Sge                             &  $81.63$ & $0.52^{+0.014}_{-0.01}$   &                   & 8, 6 \\  
SDSS~J080434.20$+$510349.2         &  $84.97$ & $0.32\pm0.03$             &                   & 8, 6 \\
SDSS~J105754.25$+$275947.5         &  $90.42$ & $0.60\pm0.20$             & $0.0436\pm0.0020$ & 9 \\     
GW~Lib                             &  $76.78$ &                           & $0.050\pm0.007$   & 10 \\
EZ~Lyn                             &  $84.66$ &                           & $0.042\pm0.014$   & 11 \\
CRTS~J122221.6$-$311525            & $109.80$ &                           & $0.036\pm0.010$   & 12 \\
SDSS~J125044.42$+$154957.4\tablefootmark{a}   &  $86.30$ & $1.7\times10^{-4}$        &                   & 13 \\      
V379~Vir\tablefootmark{a}                     &  $88.40$ & $5.3\times10^{-4}$        &                   & 13 \\
SDSSJ~151415.65$+$074446.5\tablefootmark{a}   &  $88.70$ & $5.9\times10^{-4}$        &                   & 13 \\
SMSSJ~160639.78$-$100010.7\tablefootmark{a}   &  $92.00$ &                           & $0.065\pm0.005$   & 14 \\
\hline
\end{tabular}
\tablefoot{\tablefoottext{a}{Magnetic systems that are presumably period bouncers and the detached magnetic white dwarf plus brown dwarf binary SMSSJ\,160639.78$-$100010.7 which might be a detached CV.}}
\tablebib{
(1)~\citet{2008MNRAS.388.1582L} (2) \citet{2019MNRAS.486.5535M}; (3) \citet{2006Sci...314.1578L}; (4) \citet{2018MNRAS.481.2523P}; (5) \citet{unda-sanzanaetal08-1}; (6) \citet{palaetal22-1}; (7) \citet{2023MNRAS.523.6114N}; (8) \citet{2008A&A...486..505Z}; (9) \citet{2017MNRAS.467.1024M}; \citet{2010ApJ...715L.109V}; (11) \citet{2021ApJ...918...58A}; (12) \citet{2017MNRAS.467..597N}; (13) \citet{2023A&A...676A...7M}; (14) \citet{kawkaetal21-1}
}
\end{table*}

\end{appendix}

\end{document}